\documentclass{kluwer}    

\usepackage{epsfig}

\newcommand{\arcsec}{\hbox{$^{\prime\prime}$}}
\newcommand{\farcs}{\hbox{$.\!\!^{\prime\prime}$}}
\newcommand{\arcmin}{\hbox{$^\prime$}}
\newcommand{\etal}{{\it et al.}}

\newcommand{\eg}{{\it e.g.,}}

\begin{document}                                                                                   
\begin{article}
\begin{opening}         
\title{Deep Galaxy Surveys in the 9150 \AA\ Airglow Window} 
\author{Alan \surname{Stockton}}  
\runningauthor{A. Stockton}
\runningtitle{Surveys in the 9150 \AA\ Airglow Window}
\institute{Institute for Astronomy, University of Hawaii}

\begin{abstract}
We describe the current status of two complementary programs to search
for objects with strong emission lines in a $\sim300$ \AA\ gap, centered
at 9150 \AA, in the strong airglow emission.  Both programs are being 
carried out with LRIS on the Keck II telescope.  The first of these uses
broad-band and narrow-band filter photometry to select candidates, 
followed by multi-slit spectroscopy through the same narrow-band filter 
to limit the bandpass and allow a dense packing of slits.  The second uses 
six parallel long slits to carry out a blind spectroscopic search through 
the filter isolating the 9150 \AA\ window.  The total slit area covered 
ranges from 1 to 3.5 square arcmin per pointing, depending on slit width, 
and we can obtain $3\sigma$ detections of emission lines of 
$<2\times10^{-18}$ erg cm$^{-2}$ s$^{-1}$ in a 12000 s observation with
1\farcs5 slits.

Because, for faint objects in both programs, we are most sensitive to strong 
lines with large equivalent widths, most of our detections will be restricted
to a few specific emission lines at certain discrete redshifts. One of the 
more interesting possibilities is Ly-$\alpha$ at $z\sim6.5$.  However, 
even with 12000 s integrations on the Keck II telescope, our narrow-band 
imaging does not pick up objects with emission-line fluxes $\lesssim10^{-17}$ 
erg cm$^{-2}$ s$^{-1}$.  With this limit, at $z\sim6.5$, we would
pick up only the most luminous of the $z>5$ objects discovered so far.
Our blind spectroscopic search potentially has a better chance of 
discovering such objects, but we have not yet found any definite examples 
in the limited area of the sky we have covered to date.  We discuss the 
criteria for identifying Ly-$\alpha$ emission in noisy spectra and emphasize 
how high-ionization dwarf galaxies at low redshift can be mistaken for 
Ly-$\alpha$ candidates under certain conditions.
\end{abstract}

\end{opening}           
\section{Introduction}
Our knowledge of the distant universe has grown immensely over the past
few years, thanks largely to methods that have been developed to
isolate populations of high-redshift galaxies from other faint objects.
Many of these methods depend on measuring Ly-$\alpha$ emission or
the continuum discontinuity caused by the Ly-$\alpha$ forest and the Lyman
break.  These techniques can be effective with large ground-based telescopes
up to $z\sim5$; at higher redshifts the observations are
seriously compromised by the strong OH airglow emission from the upper 
atmosphere. 

\begin{figure}
\centerline{\epsfig{file=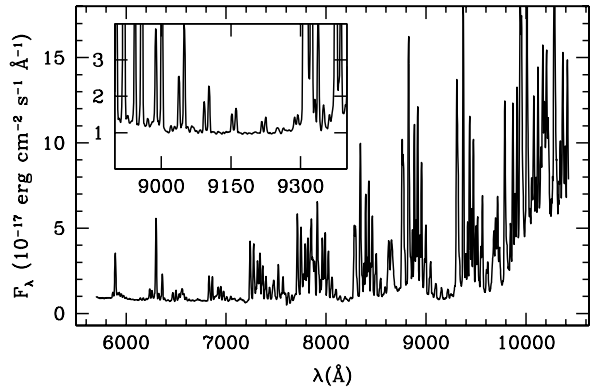,width=20pc}}
\caption{The airglow spectrum under dark conditions from Mauna Kea, obtained
with LRIS on Keck I.  The inset shows an enlargement of the region around
915 nm.}\label{skycal}
\end{figure}

As Fig.~\ref{skycal} shows, although the OH airglow bands become a major problem longward of 7240 \AA,
there are still small portions 
of the spectrum that are relatively uncontaminated, principally the 
bandpasses 200--300 \AA\ wide centered around 8150 \AA\ and 9150 \AA.
In these regions, the average sky brightness is hardly worse
than it is in the $R$ band.  We have chosen to concentrate on the
9150 \AA\ band, since it is the longest-wavelength clear region accessible
to CCD detectors.  

A combination of advances in detector development, advances in
large interference filter design and manufacture, and the availability
of 8--10~m class telescopes now
makes it possible to exploit this window for a range of
issues in extragalactic astronomy that have previously been out of reach.
Some of these form a natural bridge to topics that will be central to the
expected program for the New Generation Space Telescope.

\section{Scientific Rationale}  
The essential feature of our program is the detection of faint objects
having fairly strong emission lines with large equivalent widths. 
Most of our detections will be one of the
following:  H$\alpha$ at $z\sim0.39$; [O\,III] $\lambda\lambda4959$,5007 
at $z\sim0.83$; [O\,II] $\lambda3727$ at $z\sim1.45$; or (possibly) 
Ly-$\alpha$ at $z\sim6.5$.  Perhaps surprisingly, we can usually decide 
among these possibilities from spectra covering only a $\sim300$~\AA\ region.  

Because we are investing a lot of telescope and analysis time in each of
the fields we investigate, we have attempted to combine the emission-line
search with other programs that require deep imaging and/or spectroscopy.
In particular, we have targeted either (1) 3C radio source fields that would
place the [O\,II] $\lambda3727$ doublet within the 9150 \AA\ window, or (2)
quasar fields with $1.4<z<1.7$ in which we have old galaxy candidates 
requiring deep spectroscopy for confirmation.

There are obviously many interesting programs that can come out of
the very deep flux-limited samples we will eventually have for large
numbers of H$\alpha$ and [O\,III] emission-line sources, as well as
smaller numbers of [O\,II] sources, but we do not have space to discuss 
them here. 
Instead, we will concentrate on the possibilities for detection of
Ly-$\alpha$ emitting galaxies at $z\sim6.5$.

Early searches for so-called ``primeval galaxies''
were more-or-less predicated on a picture
of bulge and E galaxy formation in which essentially all of the star
formation takes place within a collapse time scale \cite{egg62}.
Predicted luminosities and Ly-$\alpha$ fluxes were accordingly
very high, and the lack of detections of such objects in the early
surveys (\opencite{pri87}, 1990; see also
\opencite{deP93}; \opencite{tho95}) was disillusioning,
although it did spur efforts to rethink galaxy formation mechanisms.
Views on galaxy formation have shifted radically over the past decade, partly
as a result of the recognition of the controlling role played by
dark matter and of the growing power and sophistication of n-body simulations
of the early Universe, combined with semi-analytic treatments of dissipation
and star formation (\eg\ \opencite{kau99}; \opencite{col98};
\opencite{gov98}).
According to the current view, most galaxy formation takes place via the
accumulation of baryonic matter within dark matter haloes, and the gradual
merger of these baryonic ``seeds'' as the dark matter haloes merge.  The
timing of star formation (and thus the formation of ``galaxies'' in the
usual sense) depends on the relative time scales for mergers (with resultant
shock heating of the gas) and for cooling processes (which will be inefficient
for primordial material).

In the optical/IR regime, observational
approaches towards discovering high-redshift galaxies have tended to emphasize
photometric redshift determinations based largely on absorption by the
Ly-$\alpha$ forest and the Lyman limit (\opencite{ste95},1996; 
\opencite{mad96}; \opencite{lan96}; \opencite{fer98}).
\inlinecite{tho94} discuss
the possibility of using narrow-band imaging in the IR.
However, as \inlinecite{hu98} have emphasized, there is still a place for
Ly-$\alpha$ emission searches, since there should be very little dust present
in the very first galaxy generation.  In fact, such searches should be biassed
in favor of objects approaching truly ``primeval'' galaxies.  Determining
the nature of such objects is a crucially important component for our
understanding of the early phases of galaxy formation in the Universe.

Within the past year or so, a number of objects with $z>5$ have been
claimed, most based at least partly on Ly-$\alpha$ emission (\opencite{dey98};
\opencite{hu98}, 1999; \opencite{wey98}; \opencite{spi98}; \opencite{che99}).
Some of these, at least, are quite solid and convincing.  
Our purpose is to investigate the
possibility of the developing of significant samples of objects at $z\sim6.5$.
Ly-$\alpha$ selection of objects at this redshift will give us a
unique window on the nature of early galaxy formation in the Universe.
One can use the observed Ly-$\alpha$ fluxes, somewhat cautiously, to make
estimates of star-formation rates, and therefore the contribution of such
objects to the UV ionizing continuum.  The two concerns are the effect of
the Ly-$\alpha$ forest in removing flux from the emission profile,
and the
effect of even very small amounts of dust on this resonance line.  However,
\cite{hu98} find that the equivalent widths in their filter-selected
samples (redshifts 3.4 and 4.5) were close to the maximum expected
for ionization by stars, leaving
little room for dust extinction.  In any case, one can at least determine a
lower limit to the star-formation rate.
\section{Observational Strategies}
Our filter photometric selection uses three filters: a narrow-band filter
with a center wavelength of 9148 \AA\ and a FWHM of 274 \AA\ (henceforth
$N_{915}$), an RG-850 filter combined with the CCD response cutoff 
(henceforth $Z$), and an $R$-band filter.  The $N_{915}$ and $Z$ filters have
almost the same effective central wavelengths, but the bandpass of the latter
is abouth 5 times larger, so emission-line objects can be recognized
from photometry using these two filters alone.  
The $R$ image is
useful to look for objects that are also $R$-band dropouts. We carry
out automated photometry of the $N_{915}$ image to produce a catalog of
objects detected above a given threshold.  We then do the photometry at
corresponding positions of the $Z$ and $R$ images.  Objects that are not
detected in the $Z$ filter are generally considered spurious (although
this criterion means rejecting some sources with weak, large-equivalent-width
lines).  From this 
photometry, we can construct samples of
emission-line candidates, with or without $R$-band dropout criteria.

For spectrographic confirmation, we first obtain multislit spectra of
candidates through the $N_{915}$ filter, and with moderately high dispersion.
This procedure allows us to
obtain simultaneous spectroscopy of $>100$ objects at once, but only with 
very restricted wavelength coverage. Nevertheless, we can often obtain a
firm redshift from these spectra alone: H$\alpha$ often is accompanied
by [N\,II] $\lambda6583$; in about 2/3 of the cases where we see 
[O\,III] $\lambda5007$, [O\,III] $\lambda4959$ is also within the bandpass; 
we can resolve
[O\,II] $\lambda3727$; and we should also be able to resolve the expected
profile asymmetry in Ly-$\alpha$ caused by absorption of the blueward wing
by the Ly-$\alpha$ forest. For cases that remain uncertain, we obtain
conventional wide-band multislit spectroscopy in order to pick up other
spectral features.

In order to go to fainter limits than is possible with filter photometry,
we have also attempted a blind spectroscopic search. We still limit our
bandpass with the $N_{915}$ filter, allowing us to use 6 parallel
long slits simultaneously.  The targets of interest are mostly 
$\lesssim1\arcsec$ in size, so we can adjust the slit widths to trade
between sky coverage and sensitivity without seriously compromising
spectral resolution.  We have used both 1\farcs5 and 5\arcsec\ slits in
our observations to date. For wide-bandwidth followup spectroscopy of
candidates, we use conventional multislit masks, with the slitlets oriented
nearly perpendicular to the long slits (since, especially for the wider
parallel slits, the position of an object within the width of the slit 
is uncertain).
\section{Results}
\subsection{Observing Program}
We have carried out some pilot programs with Keck II/LRIS
in the fields of 3C\,298 and
3C\,437 using imaging selection and in the fields of 3C\,280.1 and
4C\,15.55 using spectroscopic selection. The total integrations in our
three filters for the 3C\,298 and 3C\,437 fields were, respectively,
6000 s and 22800 s in $N_{915}$, 6000 s and 10800 s in $Z$, and 2700 and
12000 in $R$. While the total integrations were in all cases longer for
the 3C\,437 field, observing conditions were more variable, so the longer
integrations did not give the detectivity gain that one might have
expected.  Our 3$\sigma$ $A\!B$ magnitude 
formal limits in a 2\arcsec-diameter aperture were 25.4, 26.2, and 27.2 for the 
3C\,298 field for the $N_{915}$, $Z$, and $R$ filters, respectively, and
25.7, 26.2, and 28.0 for the 3C\,437 field for the same filters.
These correspond to detection limits for emission lines in the $N_{915}$ 
filter of
$\sim3\times10^{-17}$ erg cm$^{-2}$ s$^{-1}$ for the 3C\,298 field and
$\sim2\times10^{-17}$ erg cm$^{-2}$ s$^{-1}$ for the 3C\,437 field.  
Because our images are typically 0\farcs7 FWHM or better, we can actually
use smaller apertures and improve on these formal limits by about a 
factor of 2.  For
our spectroscopic selection images from 1\farcs5 slits, with a typical
integration of $\sim12000$ s, the emission-line limit (again, 3$\sigma$)
was $\sim1.5\times10^{-18}$ for an aperture with a diameter of 2\arcsec\ in
the spatial direction and 12 \AA\ in the dispersion direction, and
in a region free of airglow lines. For nearly
stellar objects, the limit was about twice this when we used 5\arcsec\
slits. As a reference, the emission-line flux of the galaxy SSA22-HCM1,
at $z=5.74$ \cite{hu99}, would correspond to $1.1\times10^{-17}$ 
($1.3\times10^{-17}$) erg cm$^{-2}$ s$^{-1}$ if it were seen at $z=6.5$,
assuming $q_0=0$ (0.5).

\begin{figure}
\centerline{\epsfig{file=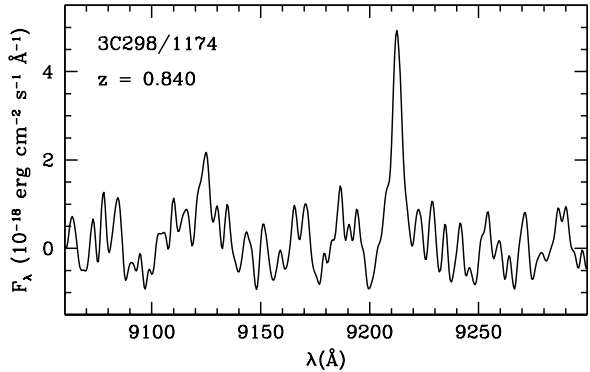,width=20pc}}
\caption{An example of a faint emission object in our 3C\,298 field.
The object has magnitude $A\!B_{9150}=26.8$ in the RG850 filter.  The two lines
visible in this spectral region are [O\,III] $\lambda\lambda4959$,5007, at
a redshift of 0.84. The [O\,III] $\lambda5007$ line has a flux of
$3\times10^{-17}$ erg cm$^{-2}$ s$^{-1}$.
}\label{s1174}
\end{figure}

We have only just begun exploring these datasets, and we have not yet found
any firm Ly-$\alpha$ detections, although we have a number of possible
candidates. Our observations cover $\sim40$ square arcmin in imaging mode and
$\sim5$ square arcmin in spectroscopic mode. Unfortunately, at this
stage we cannot place useful limits on star-formation rates in Ly-$\alpha$
emitters at $z\sim6.5$ because of the incompleteness of our spectroscopic
follow-up. Figure \ref{s1174} shows an example of a spectrum of a faint 
object found in our 3C\,298 field.
\subsection{Line Identification Pitfalls}
\begin{figure}[t]
\centerline{\epsfig{file=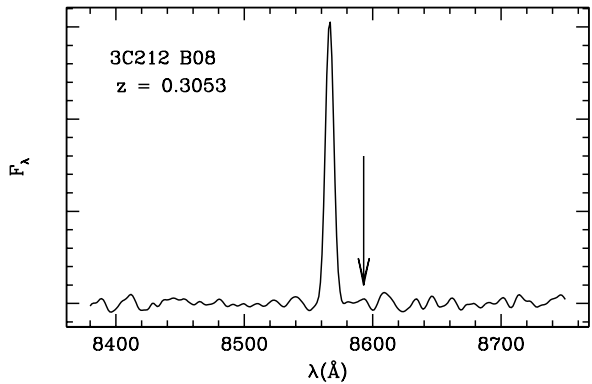,width=20pc}}
\caption{The spectrum of the region near the H$\alpha$ line in a magnitude
25 ($AB_{\rm 700 nm}$) object in the field of 3C\,212.  The arrow shows the
expected position of [N\,II] $\lambda6583$.  The identification of the
line as H$\alpha$ was based on the detection of weak He\,I $\lambda5876$
emission.  See Stockton and Ridgway (1998) for more details.
}\label{b08}
\end{figure}

Because luminous Ly-$\alpha$ emitters at very high redshift are quite rare,
an argument in favor of identifying a given line as Ly-$\alpha$ solely by
a process of elimination must always be viewed with some suspicion. 
Figure \ref{b08} shows the spectrum of a faint
object from our spectroscopy of the 3C\,212 field \cite{sto98}.
The strong emission line at 8567 \AA\ has an observed equivalent width
of 640 \AA, and no other emission lines are apparent over the observed
wavelength range from 6855 to 9477 \AA. One can eliminate most of the
obvious possibilities, and at one point we thought that the line might be
Ly-$\alpha$.  But careful measurement of the extremely weak continuum showed
no discontinuity across the line, and we finally determined that the line
must be H$\alpha$, in spite of the absence of [N\,II] $\lambda6583$, 
by finding very weak
He\,I $\lambda5876$ emission. Similar objects, though rare, are known locally;
an example is shown in Fig.~4o of \inlinecite{ter91}.  The low metallicities
and high ionizations of such objects combine to almost totally suppress 
low-ionization
metal lines. Note that if our spectral range had extended to slightly bluer
wavelengths, we would have detected very strong [O\,III] emission, which would
have resolved the issue immediately. 

We do have a fair number of cases showing only a single, moderately strong 
emission line in our 300 \AA\ bandpass. For most of these, we can eliminate
the possibility that they are
either [O\,II] $\lambda$3727 or [O\,III] $\lambda\lambda4959$,5007.
The observed equivalent widths are typically $>200$ \AA, and, if the 
emission is connected with star formation, almost the only plausible
possibilities are H$\alpha$ in a low-metallicity dwarf galaxy, H$\beta$ in 
almost any active starburst galaxy, or Ly-$\alpha$; the higher Balmer lines 
and most other lines are unlikely to have such large equivalent widths.
A fairly strong test would be to obtain good 
spectroscopy covering the region near 7000 \AA, where [O\,II] $\lambda$3727
would fall if the line in the 9150 \AA\ window is H$\beta$, and where
[O\,III] $\lambda5007$ would fall if the line is H$\alpha$.
\subsection{Future Plans}
In trying to carry out complete campaigns on our fields in a single observing
season, we have targeted only the most obvious emission-line candidates. We
plan now to do a more thorough look at each of our fields. Our imaging
data so far goes to a depth that is likely right on the edge of detecting
the brightest Ly-$\alpha$ sources at $z\sim6.5$.  We hope to be able to
double our present effective imaging integration time in these two fields.  Over the
slightly longer term, we plan to bring this program to the Subaru telescope
prime focus camera, where the field of $24\arcmin\times30\arcmin$ and the
use of deep-depletion CCDs (which will double our quantum efficiency at
9150 \AA) will give us a factor of 40 improvement in throughput.
\acknowledgements
\vspace{-2mm}
I thank Gabriela Canalizo for help with the observing,
Esther Hu for many useful discussions on
searches for high-redshift emission-line objects, and Adam Stanford,
the referee, for helpful comments. The Keck staff have been
extremely helpful; 
in particular, Bill Mason, Barb Schaefer, and Greg Wirth each found creative
solutions to various problems. I also thank the U.S. National Science 
Foundation for supporting this research under grant AST-952078.

\end{article}
\end{document}